\documentclass[review,sort&compress,times]{elsarticle}
\journal{Elsevier}
\usepackage[utf8]{inputenc}
\usepackage{lineno,color}
\usepackage{lscape}
\usepackage{amssymb,amsmath,amsthm,amsfonts}
\usepackage{graphicx}
\usepackage{graphics}
\usepackage[usenames,dvipsnames,svgnames,table]{xcolor}
\usepackage{hyperref}

\newsavebox{\bigleftbox}
\hypersetup{
 colorlinks,
 citecolor=Blue,
 linkcolor=Blue,
 urlcolor=Blue
}

\usepackage[symbol]{footmisc}

\begin{document}

\begin{frontmatter}

\title{Dynamics and Structural Transformations of Carbon Onion-Like under High-Velocity Impacts}

\author[ifd]{M. L. Pereira J\'unior}
\author[ifd]{W. F. da Cunha}
\author[ene]{R. T. de Sousa Junior}
\author[ene]{G. D. Amvame Nze}
\author[ene]{D. S. Galv\~ao}
\author[ifd]{L. A. Ribeiro J\'unior\footnote[1]{Corresponding author. E-mail: ribeirojr@unb.br (Luiz Ant\^onio Ribeiro J\'unior). Tel: +55 61 3107-7700.} }

\address[ifd]{Institute of Physics, University of Bras\'{i}lia, Bras\'{i}lia 70919-970, Brazil.}
\address[ene]{Department of Electrical Engineering, University of Bras\'{i}lia 70919-970, Brazil.}
\address[unicamp1]{Applied Physics Department, University of Campinas, Campinas, S\~ao Paulo, Brazil.}
\address[unicamp2]{Center for Computing in Engineering and Sciences, University of Campinas, Campinas, S\~ao Paulo, Brazil.}


\begin{abstract}
Carbon nano-onions (CNO) are multi-shell fullerenes. In the present work, we used fully atomistic reactive (ReaxFF) molecular dynamics simulations to study the dynamics and structural transformations of CNO structures under high-velocity impacts against a fixed and rigid substrate. We considered single and multi-shell CNO (up to six shells) and at different impact velocities (from 2 up to 7 Km/s). Our results indicated three regimes formed after the CNO impact: slightly deformed CNO (quasi-elastic collision, below 2.0 Km/s), collapsed CNO (inelastic collisions, between 3.0 and 5.0 Km/s) forming a diamondoid-like core, and fragmented CNO yielding linear atomic carbon chains (above 5.0 Km/s). We also discussed the dynamical reconfiguration of carbon-carbon bonds during the collision process. The impact of CNO against the substrate yielded $sp^3$-like bond types for all the used initial velocities. At intermediate velocities (between 3.0 and 5.0 Km/s), the inelastic collision forms diamondoid-like cores by converting a substantial quantity of $sp^2$ bonds into $sp^3$ ones. In the high velocities regime, the total number of $sp^1$, $sp^2$, and $sp^3$ bonds tend to be similar.   
\end{abstract}

\end{frontmatter}


\section{Introduction}
Different types of carbon-based materials have been the subject of great interest from the scientific community due to their unique properties and potential applications \cite{zhang2009carbon,gopinath2021environmental}. As a result, a substantially increased understanding of systems as diverse as conjugated polymers \cite{kraft1998electroluminescent}, carbon nanotubes \cite{iijima1991helical}, graphene nanoribbons \cite{han2007energy}, and fullerenes \cite{kroto1985c} has been achieved. Studies on their structural and electronic properties from both experimental \cite{zhang2007electrical,keru2014review,shuttle2008experimental,zhang2013experimentally,dutta2010novel} and theoretical \cite{ouyang2009theoretical,abergel2010properties,bauernschmitt1998experiment,greaney1991production} points of view stand out. These features are considered crucial for improving electrochemical energy storage \cite{yao2015carbon,lv2016graphene} and conversion \cite{trogadas2014carbon,wang2014recent} processes in the electronics field. 

The fullerenes discovery in the early 80s \cite{iijima1991helical} triggered the synthesis of a novel carbon allotrope based on C$_{60}$ in the 90s. Ugarte reported multilayered fullerenes by working with curved graphitic networks, today known as Carbon Nano-Onions (CNO) \cite{ugarte1992curling}. These systems exhibit a unique combination of structural and electronic properties \cite{zeiger2016carbon}. CNO are quasi-spherical nanostructures composed of multiple enclosed concentric fullerene shells (in a close analogy to the layers of an actual onion, see Figure \ref{fig:systems}) \cite{zeiger2016carbon}. Diamond-like core structures have also been observed \cite{banhart1996carbon}. These systems present high electrical conductivity levels and a large surface area when contrasted with other nanomaterials \cite{zeiger2016carbon}. Their relatively easy synthesis provides a level of controllability that allows efficient functionalization procedures \cite{bartelmess2014carbon,qin1996onion}. Such properties have made CNO good candidates for a wide range of applications, such as biological imaging and sensing \cite{ghosh2011carbon,sonkar2012carbon}, environmental remediation \cite{luszczyn2010small,seymour2012characterization}, electronics (as batteries, capacitors, fuel cells, and terahertz-shielding devices) \cite{bartelmess2014carbon,han2011preparation}, catalysts \cite{su2007oxidative}, tribology \cite{cabioc2002structure}, optical limiting \cite{koudoumas2002onion}, and molecular junctions in STM \cite{sek2013stm}.  

It is well-known that materials can have their properties considerably altered when subjected to extreme conditions \cite{oliveira2018hardening}. The changes are highly dependent on their structural nature and the conditions applied to them. The high-velocity impact of nanostructures can give rise to a myriad of structural deformations, as it was demonstrated for the cases of nanotubes \cite{ozden2014unzipping,armani2021high} and nanoscrolls \cite{woellner2018structural,de2016carbon}. This technique has even been considered a viable mechanism for synthesizing different materials with a whole different set of properties \cite{oliveira2018hardening,woellner2018structural}. To this date, however, a similar theoretical investigation of the effects of the high-velocity impact of CNO is still lacking.

In this work, we carried out molecular dynamics simulations to investigate the high-velocity impact of CNO into a rigid substrate. To address these collision processes, we study in detail the structural, energetic, and stress aspects of these CNO/substrate impacts.  

\section{Methodology}

\begin{figure*}[!b]
	\centering
	\includegraphics[width=\linewidth]{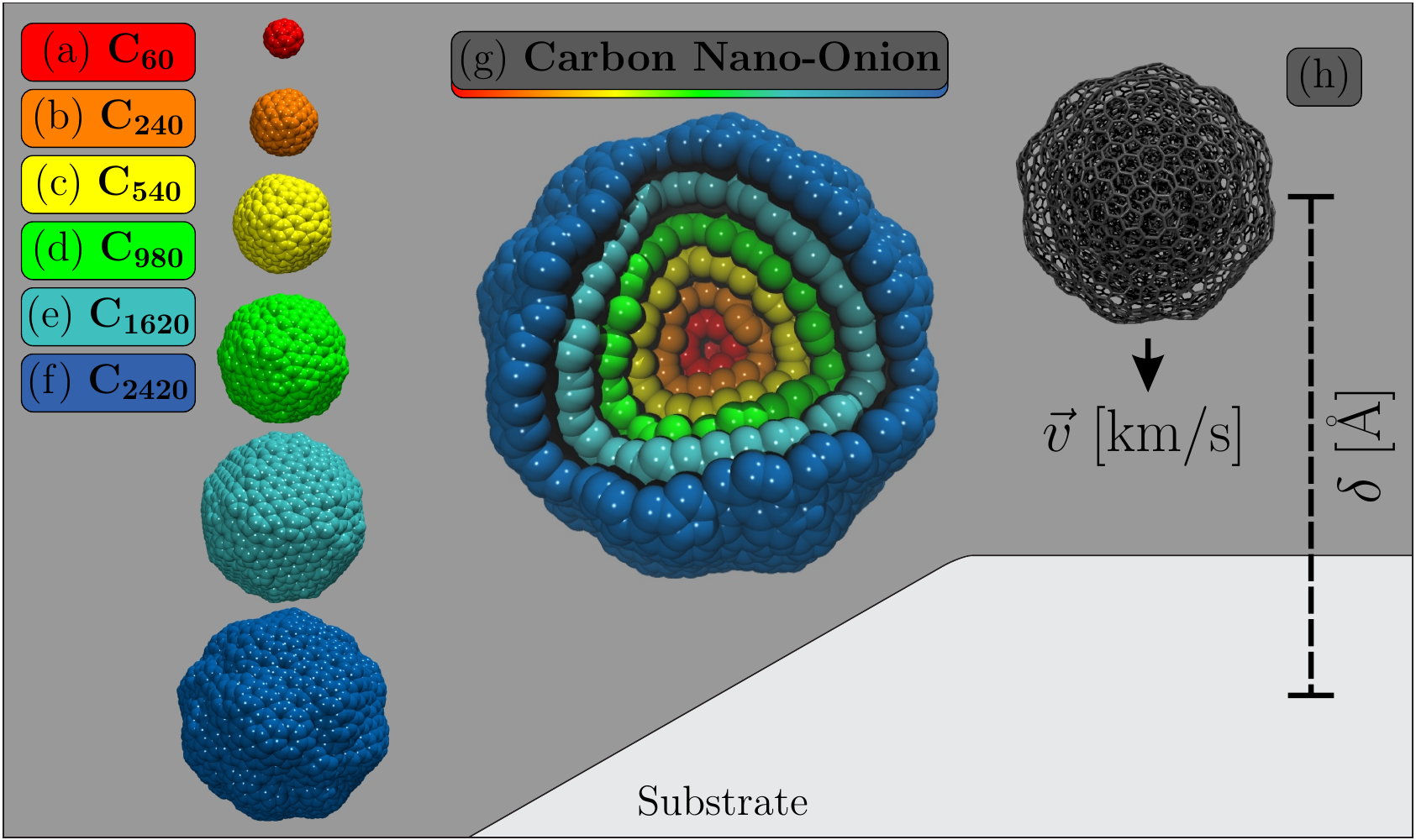}
	\caption{Schematic representation of the structural models of the carbon nano-onions used here. (a-f) illustrate the single structures (shells) that compose the full CNO, in which the number of carbon atoms varies from 60 up to 2420. The biggest multi-shell CNO has 5860 atoms. (g) depicts the composition of the CNO. (h) shows our initial  simulation setup for the CNO/substrate system in which $\delta=50$ \r{A}. This CNO structure was created based on reference \cite{qi2018adsorption}.}
	\label{fig:systems}
\end{figure*}

We carried out fully atomistic reactive (ReaxFF) Molecular Dynamics simulations \cite{vanduin_JPCA,ashraf2017extension} to investigate the impact of multi-shell CNO and their isolated shells (C$_N$ fullerenes, where $N$ stands for the number of atoms). Such simulations were conducted using the ReaxFF potential as implemented in the Large-scale Atomic/Molecular Massively Parallel Simulator (LAMMPS) \cite{plimpton_JCP}. The nanostructures obtained from reference \cite{qi2018adsorption} were edited to isolate the shells and smaller CNO using the Visual Molecular Dynamics (VMD) software \cite{humphrey1996vmd}. 

The typical CNO structure considered in this work is composed of an innermost shell of a C$_{60}$ enclosed by a varying number --- from two up to seven --- of concentric buckyballs. In Figure \ref{fig:systems} we present a schematic representation of a CNO with six shells, in which the number of carbon atoms varies from 60 up to 2420 atoms. The biggest multi-shell CNO has 5860 atoms. The CNO are formed by the composition of the different shells. The number of carbon atoms is included in the notation describing each CNO. As mentioned before, CNO$_N$ corresponds to a CNO composed of a particular number of shells. $N$ denotes the sum of the carbon atoms in these shells, from the smallest to the biggest fullerenes following the sequence shown in Figure \ref{fig:systems} (e.g., CNO$_{300}$ is a CNO composed by two shells: C$_{60}$ and C$_{240}$). Our goal is to investigate the dynamics of the systems collision at different velocities with a fixed and rigid substrate. The CNO center of mass is initially placed 50 \r{A} from the substrate. 

The rigid substrate is considered using a fixed (time-independent) 12-6 Lennard-Jones potential. Such potential was parameterized using 10 \r{A} cutoff distance, 0.07 Kcal/mol of interaction strength, and 3.55 \r{A} distance between the particles for their interaction. We adopted a non-periodic unit cell with such repulsive potential between the nanostructures and the model substrate. We made use of a microcanonical (NVT) ensemble --- to thermally equilibrate all structures at 300 K before the collision --- and a microcanonical (NVE) ensemble \cite{hoover1985canonical} for simulating the impact, considering the fixed time step of 0.1 fs. The simulations were carried out during 10-50 ps, depending on the shooting velocity.

\section{Results}

The collision process of isolated shells and multi-shell CNO on a rigid substrate were investigated here. Each shell is a buckyball composed of between 60 to 2420 atoms. The multi-shell CNO considered in this work are composed of two up to seven shells, thus having from 300 up to 5860 carbon atoms, totaling 77 cases investigated, subjected to shooting velocities varying from 1.0 up to 7.0 Km/s.  

The investigated systems were the following: As isolated shells, we considered C$_{60}$, C$_{240}$, C$_{540}$, C$_{980}$, C$_{1620}$ and C$_{2420}$. Also, the following multi-shell CNO were studied: CNO$_{300}$ (two shells: C$_{60}$ + C$_{240}$), CNO$_{840}$ (three shells: CNO$_{300}$ + C$_{540}$), CNO$_{1820}$ (four shells: CNO$_{840}$ + C$_{940}$), CNO$_{3340}$ (five shells: CNO$_{1820}$ + C$_{1620}$), CNO$_{5860}$ (six shells: CNO$_{3340}$ +  C$_{2420}$).     

\begin{figure*}[!t]
	\centering
	\includegraphics[width=\linewidth]{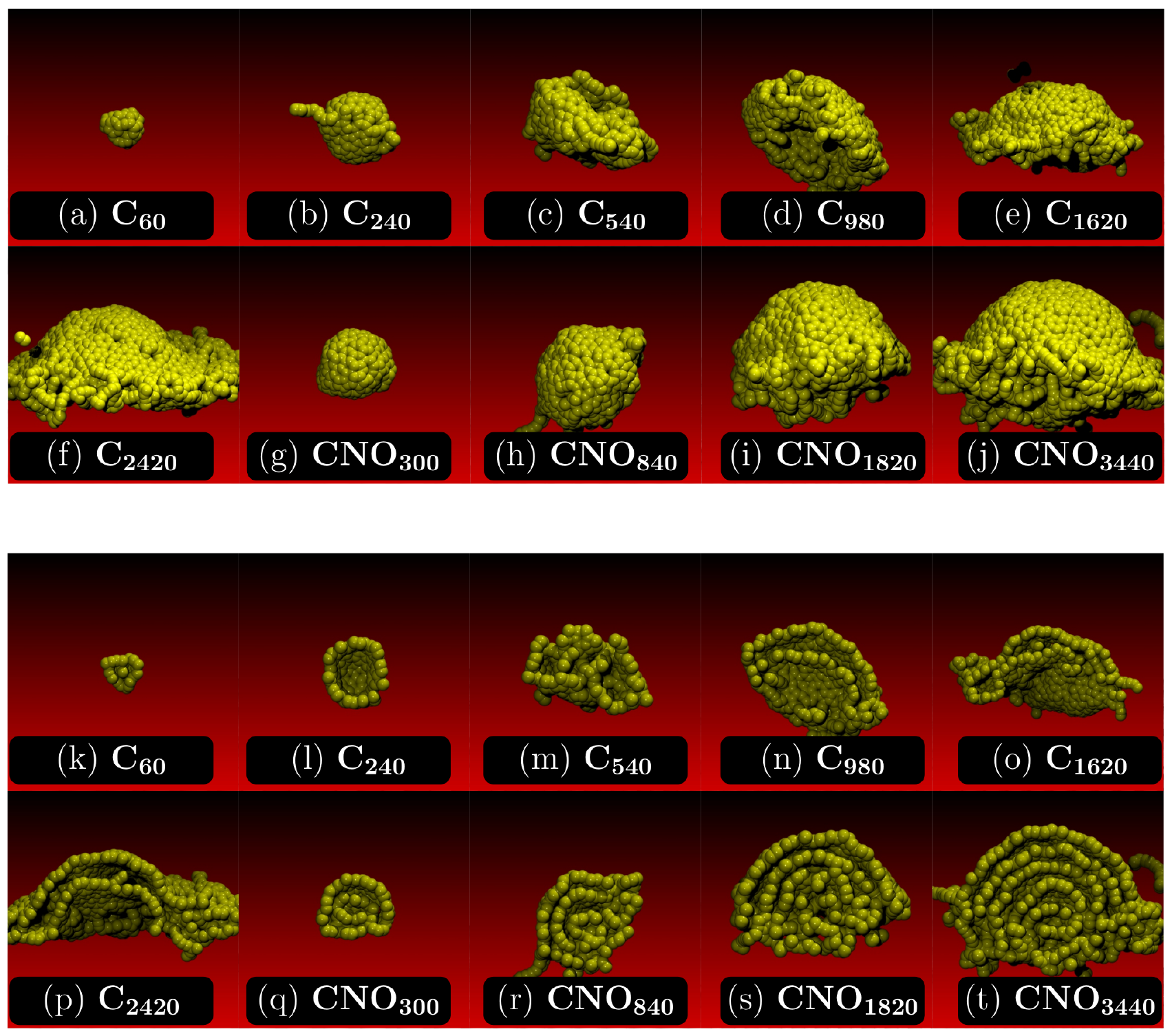}
	\caption{Representative MD snapshots for (a-f) the isolated shells and (g-j) for the CNO after their impacts. Here, CNO$_N$ corresponds to a CNO composed of a particular number of shells. $N$ denotes the sum of the carbon atoms in these shells, from the smallest to the largest fullerenes following the sequence shown in Figure \ref{fig:systems} (e.g., CNO$_{300}$ is a CNO composed by two shells: C$_{60}$ and C$_{240}$). In these cases, the shooting velocity is 5.0 Km/s. (a-j) and (k-t) show front views and the corresponding cross-section views of the isolated shells, respectively.}
	\label{fig:diff_sys}
\end{figure*}

Figure \ref{fig:diff_sys} shows the representative MD snapshots (taken at 4.0 ps of simulation) for the impact of the fullerenes raging from C$_{60}$ up to CNO$_{3440}$, shooting velocity of 5.0 Km/s. The collisions occurred before this time for all cases. The figure is divided into two panels. The upper panel shows a general view of the structure resulting from the collision. The lower one presents a cross-section view of the shells and multi-shell CNO to indicate the internal structural changes resulting from the impact. As a general trend, it is possible to observe that isolated shells possess very distinct deformation pattern dynamics. This trend is not true for the multi-shell CNO since the larger systems are less deformable than smaller ones.  

\begin{figure*}[!t]
\centering
\includegraphics[width=\linewidth]{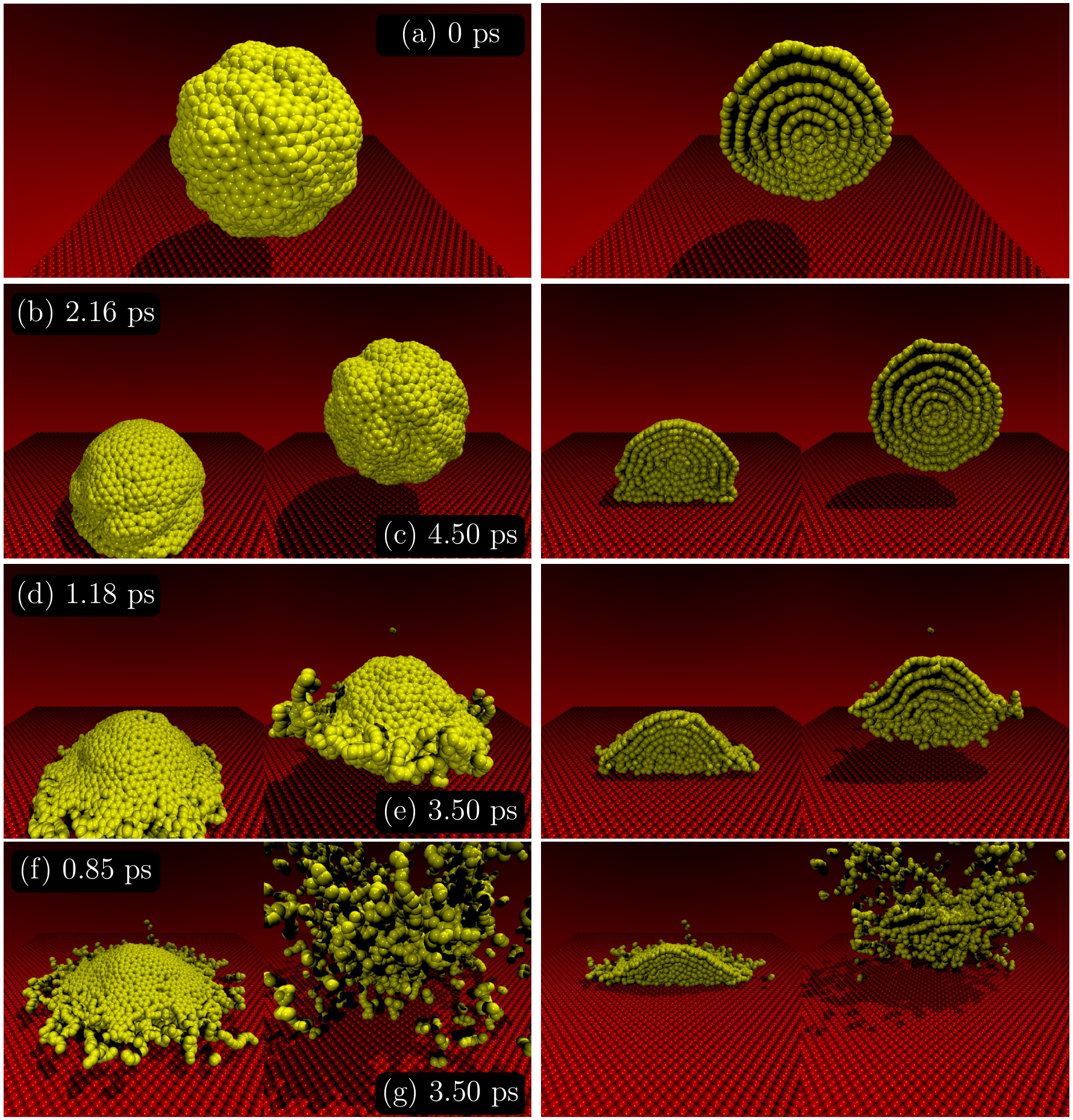}
\caption{Representative MD snapshots for the multi-shell CNO (CNO$_{5860}$) impact at different shooting velocities: 2.0 Km/s (top panels), 5.0 Km/s (middle panels), 7.0 Km/s and (bottom panels). The left and right panels illustrate the front views and their corresponding cross-section views, respectively. The Supplementary Material shows the videos of the simulations for all the cases illustrated in this figure.}
\label{fig:CNO5860}
\end{figure*}

We now discuss the case of the biggest multi-shell CNO considered in this work: CNO$_{5860}$. In Figure \ref{fig:CNO5860}, we compare the collisions results for 2.0, 5.0, and 7.0 Km/s shooting velocity values. The general and cross-section views are presented in the left and right panels, respectively. The chosen velocities are representative of three observed regimes. Supplementary Material shows the videos of the simulations for all the cases illustrated in this figure. At low velocities, up to 2 Km/s, we observed quasi-elastic collisions. This trend is clear comparing the original structure (Figure 3(a)) to the resulting ones (Figures \ref{fig:CNO5860}(b) and \ref{fig:CNO5860}(c)). At intermediate velocities, between 3.0 and 5.0 Km/s (see Figures \ref{fig:CNO5860}(d) and \ref{fig:CNO5860}(e)), inelastic collision occur, as evidenced by the substantial structural deformations. Finally, further increasing the velocities resulted in collapsed/fragmented structures, as can be seen in Figures 3(f) and 3(g). These regimes are similar to the ones reported for the cases of nanotubes \cite{ozden2014unzipping,armani2021high} and nanoscrolls \cite{woellner2018structural,de2016carbon}.

\begin{figure*}[!t]
\centering
\includegraphics[width=\linewidth]{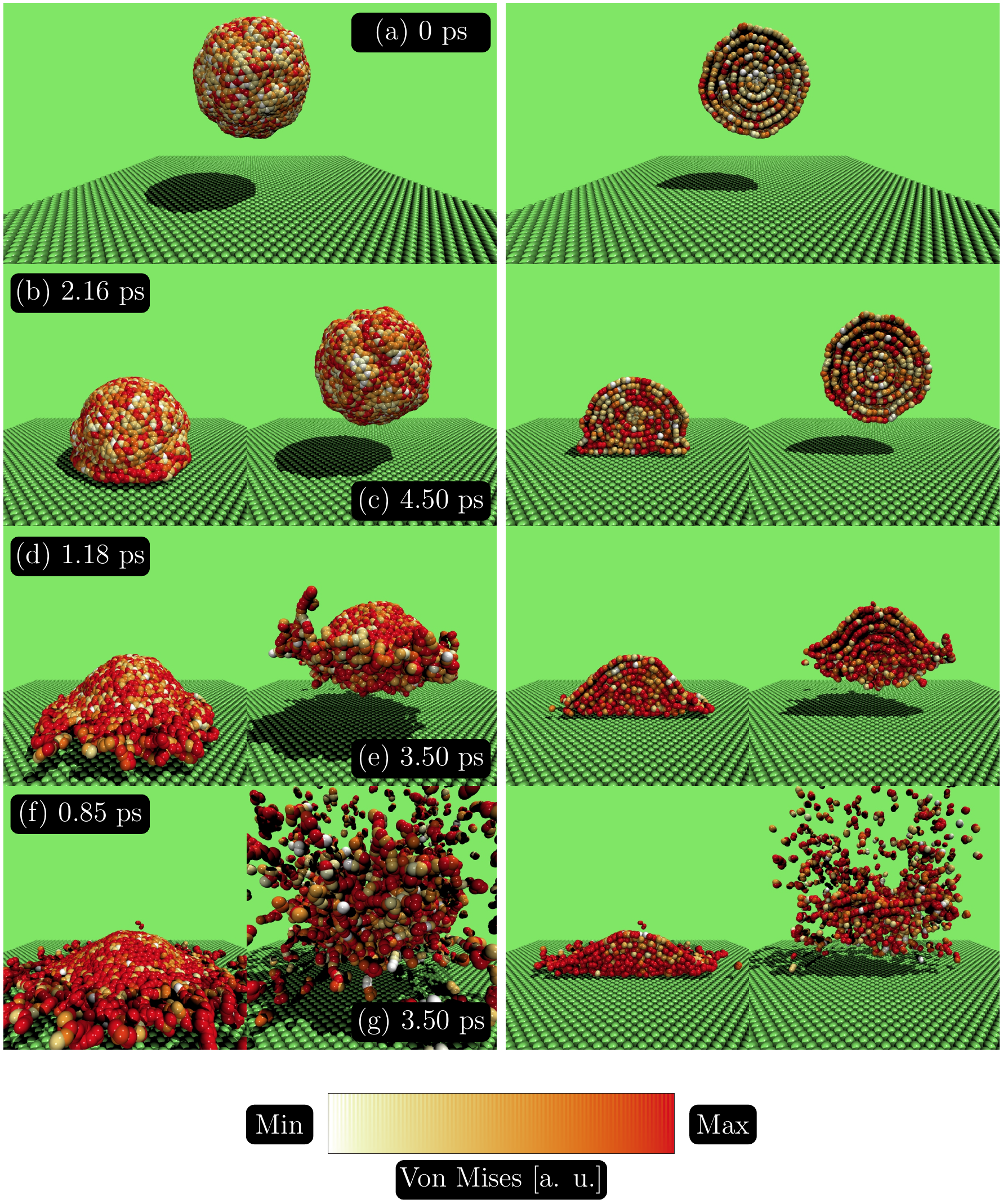}
\caption{von Mises stress distribution as a function of the simulation time for the multi-shell CNO (CNO$_{5860}$) impact at different shooting velocities: 2.0 Km/s (top panels), 5.0 Km/s (middle panels), 7.0 Km/s and (bottom panels). The left and right panels illustrate the front views and their corresponding cross-section views, respectively.}
\label{fig:vm}
\end{figure*}

To gain further insights into the CNO collision mechanisms, we analyzed the stress per atom pattern (the von Mises Stress) during the impact (Figure \ref{fig:vm}), for the cases discussed in Figure \ref{fig:CNO5860} (the CNO$_{5860}$ collision using 2.0 (Figures \ref{fig:vm}(b) and \ref{fig:vm}(c)), 5.0 (Figures \ref{fig:vm}(d) and \ref{fig:vm}(e)), and 7.0 Km/s (Figures \ref{fig:vm}(f) and \ref{fig:vm}(g)) as shooting velocities). In the color scheme adopted for this figure, white and red colors denote regions with null and maximum stress values per atom, respectively. The Von Mises stress is calculated accordingly to the procedure described in the reference \cite{felix2020mechanical}. In Figure \ref{fig:vm}(a), one can observe that the optimized structures of the model CNO used here are intrinsically stressed, as expected. At low velocities, up to 2 Km/s (Figures \ref{fig:vm}(b) and \ref{fig:vm}(c)), the stress per atom pattern does not change substantially after the collision. In this way, the quasi-elastic collisions can preserve regions with null stress (white regions). For the inelastic collisions, the CNO structure accumulates a significant amount of stress. Due to its topology, the stress accumulation is almost equally distributed over its surface after the collision, as illustrated in Figures \ref{fig:vm}(d-g). This trend for the distribution of the accumulated stress revealed that our model CNO does not present a most probable fracture point (or region). In carbon nanotubes, for instance, the stress accumulation during the impact can occur at the edge or in the center of the nanostructure yielding bilayer graphene or a graphene membrane, respectively \cite{armani2021high}.   

\begin{figure*}[!t]
\centering
\includegraphics[width=\linewidth]{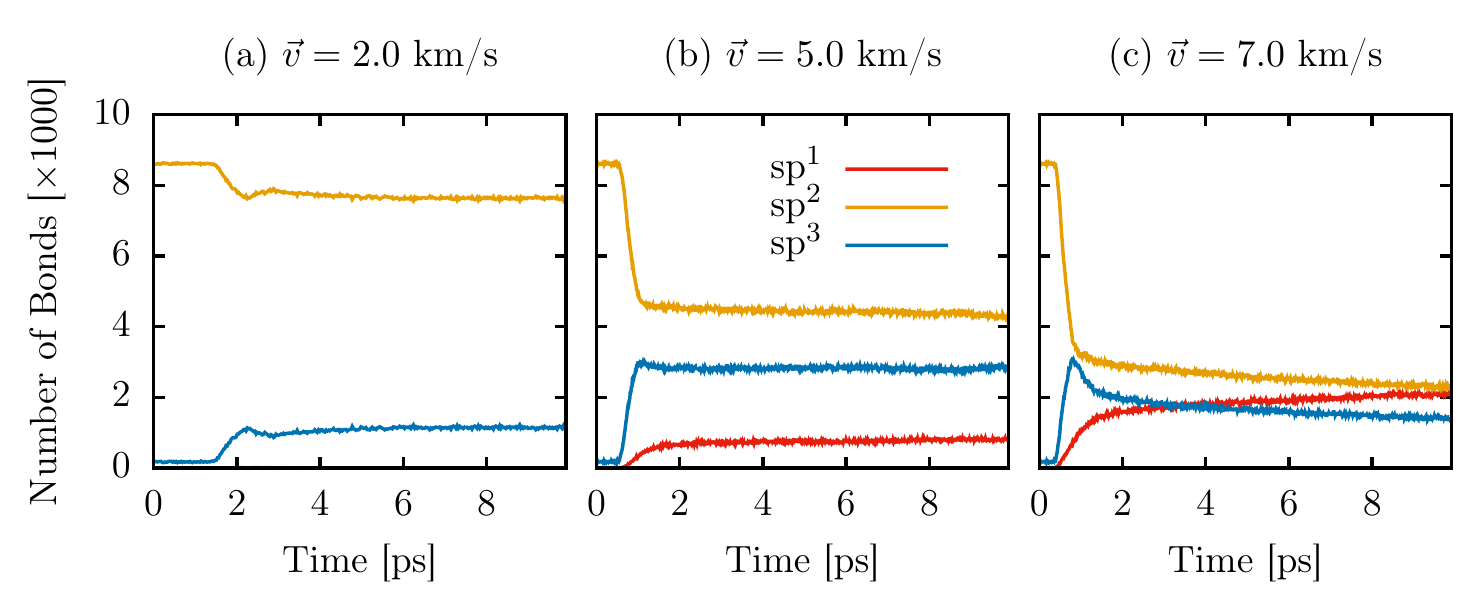}
\caption{Time evolution for the total number of different carbon-carbon bond types of the CNO$_{5860}$ system during the collision process.}
\label{fig:numbonds}
\end{figure*}         

In Figure \ref{fig:numbonds} we present the time evolution for the total number of different carbon-carbon bond types present in the CNO$_{5860}$ structure. Figures \ref{fig:numbonds}(a), \ref{fig:numbonds}(b), and \ref{fig:numbonds}(c) shows the total number of bonds ($sp^1$, $sp^2$, and $sp^3$) for the impact of the nanostructure with initial velocities of 2.0, 5.0, and 7.0 Km/s, respectively. In Figure \ref{fig:numbonds}(a), one can observe that the initial  CNO$_{5860}$ configuration only presents $sp^2$-like carbon-carbon bonds, as expected. At low velocities (up to 2 Km/s), almost 1000 bonds were converted from $sp^2$-like to $sp^3$-like ones after the quasi-elastic collision, as shown in Figure \ref{fig:numbonds}(a). The coexistence of $sp^2$ and $sp^3$ bond types remained until the end of the simulation, but with a predominance of $sp^2$ bonds, indicating only small CNO$_{5860}$ structural deformations for low velocities. In this figure, a small number of $sp^1$ bonds were also observed due to the formation of linear atomic carbon chains (LACS) (see Figure \ref{fig:CNO5860}(e)). For the high velocities regime (Figure \ref{fig:numbonds}(c)), the total number of $sp^1$, $sp^2$, and $sp^3$ bonds tend to be similar, but with a substantial decrease in the quantity of $sp^1$ and $sp^2$ bonds with relation to the two previous cases. The total number of $sp^1$ bonds increased due to a higher degree of fragmentation of the CNO$_{5860}$ and subsequent LACS formation (see Figure \ref{fig:CNO5860}(g)).   

\begin{figure*}[!t]
\centering
\includegraphics[width=\linewidth]{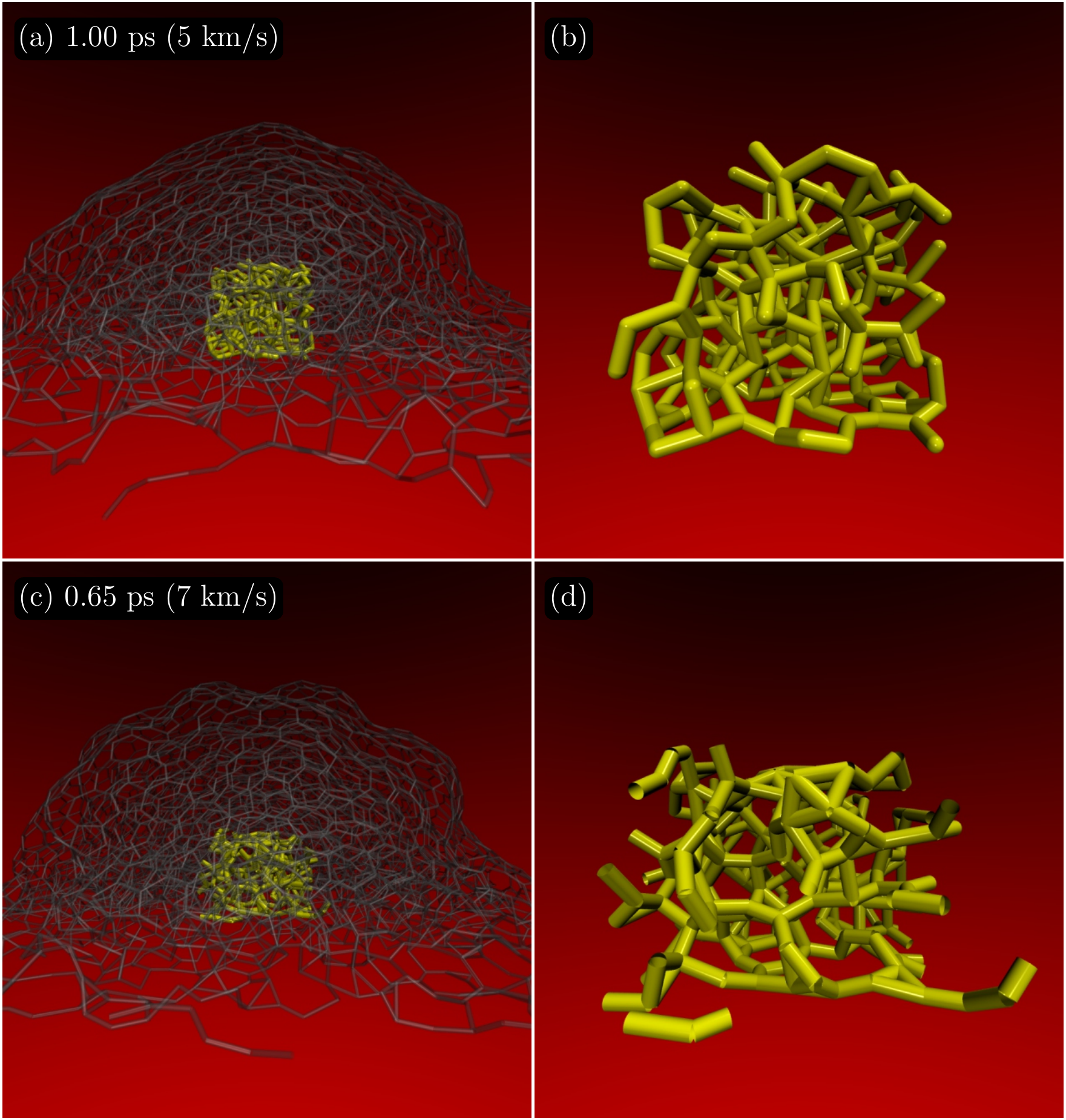}
\caption{Representative MD snapshots illustrating the formation of diamondoids after the CNO$_{5860}$ collision at (a-b) 5 Km/s and (c-d) 7 Km/s. For clarity, the yellow bonds illustrate the diamondoid-like cores, and their zoomed view is presented in the right panels (b-d).}
\label{fig:diamondoids}
\end{figure*}

One striking result was the observation of the formation of a diamondoid-like core depending on the impact velocity. In Figures \ref{fig:diamondoids}(a,c), we illustrate representative MD snapshots showing these cores for the shooting velocities of 5 and 7 Km/s. For the sake of clarity, the yellow bonds denote the diamondoid-like core, and their zoomed view is presented in the right panels, Figures \ref{fig:diamondoids}(b,d). Interestingly, all the CNO shells contributed to the diamondoid-like core formation. This behavior is a relevant result. It shows that under certain thermodynamic and/or internal pressure a diamondoid-like core can be formed in CNO. Although not always present in the experimentally realized CNO, some CNO with a diamondoid-like core has been experimentally observed \cite{banhart1996carbon}.   

\begin{figure}[!t]
\centering
\includegraphics[width=\linewidth]{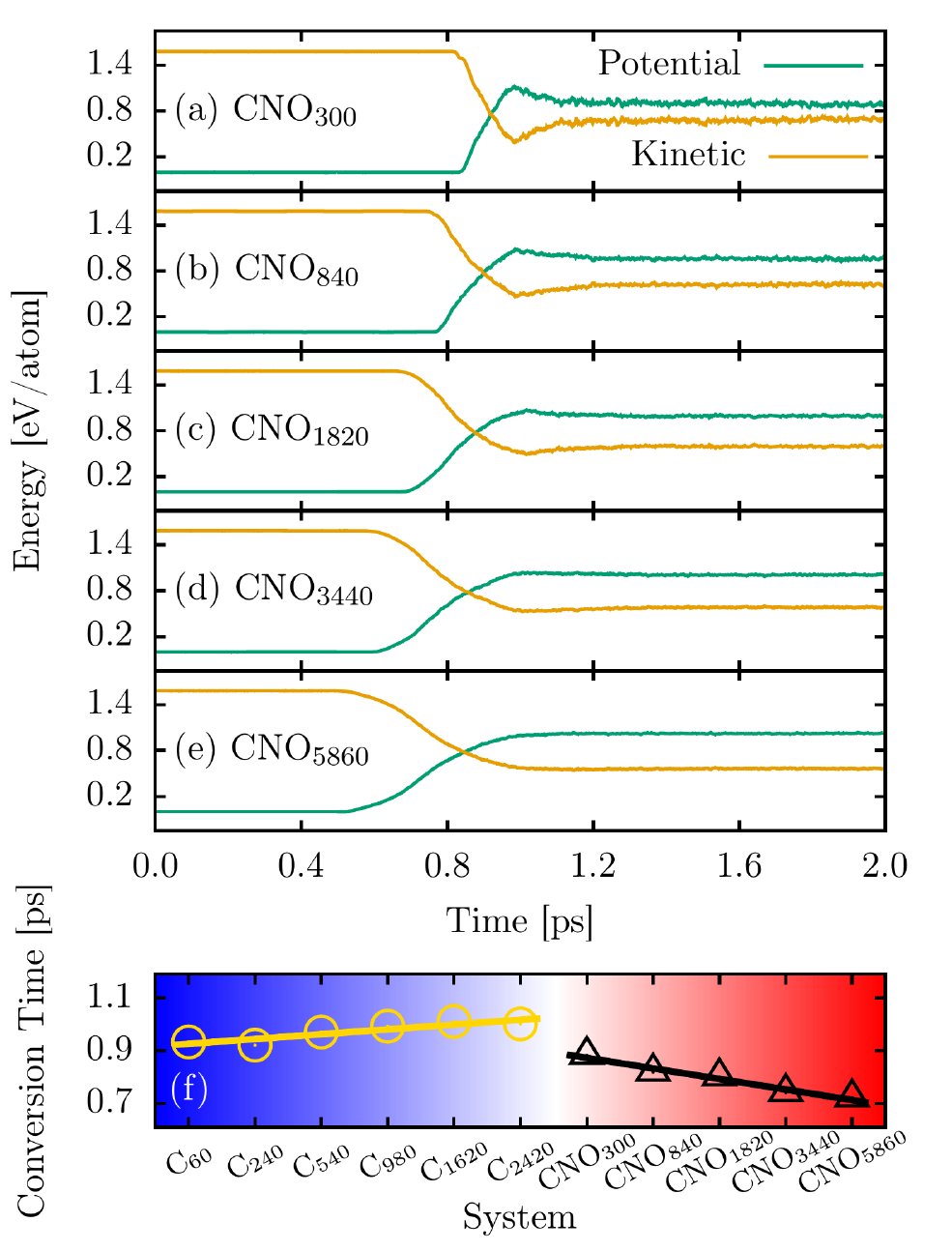}
\caption{Kinetic and potential energy values as a function of the simulation time during the CNO collision, for the velocity case of 5 Km/s. At the bottom is also indicated compression time for the different structures.}
\label{fig:energies}
\end{figure}

Finally, in Figures \ref{fig:energies}(a-e), we present the kinetic and potential energy profiles as a function of the simulation time for the multi-shell CNO structures using 5 Km/s of shooting velocity as a representative case. Figure \ref{fig:energies}(f) shows the time required for converting kinetic into potential energy for the C$_N$ (yellow line) and CNO$_N$ (black line) cases. The total potential energy was shifted to zero by subtracting the CNO ground state energy. In this sense, we are discussing the potential energy gain coming from the conversion of kinetic energy. Figure \ref{fig:energies}(f) presents the time required for the multi-shell CNO investigated here to convert their kinetic to potential energy (i.e., the conversion time during deformation). The CNO degree of deformability tends to increase when their size (i.e., the number of shells) increases. The CNO$_{300}$ and CNO$_{840}$ systems present an abrupt conversion between kinetic and potential energy (see Figures \ref{fig:energies}(a) and \ref{fig:energies}(b), respectively), that is related to the small size of the systems. For the CNO$_{1820}$, CNO$_{3440}$, and CNO$_{5860}$ cases (see Figures \ref{fig:energies}(c), \ref{fig:energies}(d), and \ref{fig:energies}(e), respectively), the conversion between these energies occurs smoothly, despite their smaller deformability when contrasted to the CNO$_{300}$ and CNO$_{840}$ cases. The big CNO (Figures \ref{fig:energies}(a) and \ref{fig:energies}(b)) interact with the substrate earlier than the small ones (Figures \ref{fig:energies}(c-e)). Although the deformation levels are different between these CNO models, the amount of energy converted is independent of the system mass. We can see that the amount of energy converted was approximately 0.7 eV/atom for all cases.  In the conversion process, the kinetic energy dropped from 1.6 to 0.7 eV/atom, while the potential energy acquired by the CNO after the collision increased from 0 to 0.9 eV/atom. Figure \ref{fig:energies}(f) shows that the conversion time has a linear trend for both C$_N$ and CNO$_N$ cases. In the C$_N$ collision, the higher is the inner shell volume the greater is the conversion time, once the deformation degree of C$_N$ shells increases with their volume. The CNO$_N$ cases, in turn, presented an inverse trend. The higher is the system mass the smaller is the conversion time. The CNO$_N$ stiffness increases with the number of shells, which makes the conversion between kinetic and potential energies faster.      
 
\section{Conclusions}

In summary, we studied the dynamics and structural transformations of carbon onion-like structures under high-velocity impacts within the framework of fully atomistic reactive molecular dynamics simulations. We considered single and multi-shell nano-onions (up to six shells) and at different impact velocities (from 2 up to 7 Km/s) against a fixed and rigid substrate.

Our results showed the existence of three different regimes, depending on the impact velocity: slightly deformed CNO (quasi-elastic collision), collapsed CNO (inelastic collisions) forming diamondoid-like cores, and fragmented CNO yielding LACS in a gas phase of carbon atoms. CNO tend to equally distribute through their surface the amount of accumulated stress during the collision. 

The impact of CNO against the substrate yielded $sp^3$-like bond types for all the used initial velocities. At intermediate velocities (between 3.0 and 5.0 Km/s), the inelastic collision forms diamondoid-like cores by converting a substantial quantity of $sp^2$ bonds into $sp^3$ ones. $sp^1$ bonds were also observed due to the formation of LACS. In the high velocities regime, the total number of $sp^1$, $sp^2$, and $sp^3$ bonds tend to be similar, around 2000. After the CNO impact and its subsequent fragmentation, the number of $sp^1$ bonds increases until the end of the simulation due to the formation of LACS.  

\section*{Acknowledgments}
The authors gratefully acknowledge the financial support from Brazilian research agencies CNPq, FAPESP, and FAP-DF. M.L.P.J gratefully acknowledges the financial support from CAPES grant 88882.383674/2019-01. D.S.G. thanks the Center for Computing in Engineering and Sciences at Unicamp for financial support through the FAPESP/CEPID Grants \#2013/08293-7 and \#2018/11352-7. L.A.R.J acknowledges the financial support from a Brazilian Research Council FAP-DF and CNPq grants $00193.0000248/2019-32$ and $302236/2018-0$, respectively. L.A.R.J acknowledges CENAPAD-SP for providing the computational facilities. L.A.R.J. gratefully acknowledges the financial support from IFD/UnB (Edital $01/2020$) grant $23106.090790/2020-86$. The authors acknowledge the National Laboratory for Scientific Computing (LNCC / MCTI, Brazil) for providing HPC resources of the SDumont supercomputer, which have contributed to the research results reported within this paper. URL: \url{http://sdumont.lncc.br}.

\bibliographystyle{model1-num-names}
\bibliography{references.bib}

\end{document}